\documentclass[11pt,letterpaper]{sa-paper}

\usepackage[backref,
            natbib
            ,style = numeric-comp
            ,maxnames = 2
            ,backend = bibtex
            ,sorting=none
            ]{biblatex}

\usepackage[T1]{fontenc}  

\usepackage[utf8]{inputenc}  
\usepackage{csquotes}
\usepackage[english]{babel}  

\usepackage{amsmath%
            ,amssymb%
            ,amsfonts%
            ,amsthm
           }  
\usepackage{mathrsfs}  
\usepackage{bm}  

\usepackage[active]{srcltx}  
\usepackage{extarrows}  

\usepackage[table,dvipsnames*,svgnames]{xcolor}
\usepackage{graphicx}

\usepackage{newtxtext, newtxmath} 

\usepackage{algorithm2e}

\newcommand{\prd}[2]{\frac{\partial #1}{\partial #2}}

\newcommand{\vev}[1]{\langle #1 \rangle}

\newcommand{\R}{\mathbb{R}}

\newcommand{\Lone}{L\textsubscript{1}\xspace}

\DeclareMathOperator{\sign}{sgn}
\DeclareMathOperator*\argmin{argmin}

\bibliography{genomics.bib}

\title{\textbf{Accurate Genomic Prediction Of Human Height}}

\author[1]{Louis Lello
}
\author[1]{Steven G. Avery
}
\author[1,3,5]{Laurent Tellier
}
\author[2]{Ana I. Vazquez
}
\author[2,4]{Gustavo de los Campos
}
\author[1,3]{Stephen D.H. Hsu
}

\affil[1]{%
\textit{Michigan State University\protect\\
Department of Physics and Astronomy\protect\\
East Lansing, MI 48824}\vspace*{0.25cm}}

\affil[2]{%
\textit{Michigan State University\protect\\
Department of Epidemiology and Biostatistics\protect\\
East Lansing, MI 48824}\vspace*{0.25cm}}

\affil[3]{%
\textit{Cognitive Genomics Laboratory, BGI}\vspace*{0.25cm}
}

\affil[4]{%
\textit{Michigan State University\protect\\
Department of Statistics and Probability\protect\\
East Lansing, MI 48824}\vspace*{0.25cm}} 

\affil[5]{%
\textit{University of Copenhagen\protect\\
Department of Biology, Functional Genetics\protect\\
Copenhagen, DK}}

\date{}

\keywords{GWAS, Quantitative Genomics}
\pdfsubject{Genomic Prediction}
\pdfauthorlist{Lello, Avery, Tellier, Vasquez, de los Campos, Hsu}

\begin{document}
\maketitle
\thispagestyle{empty} 
\begin{abstract}
  We construct genomic predictors for heritable and
  extremely complex human quantitative traits (height, heel bone density, and educational attainment)
  using modern methods in high dimensional statistics (i.e., machine
  learning). Replication tests show that these predictors capture, respectively, $\sim$40, 20, and 9 percent of total
  variance for the three traits. For example, predicted heights correlate
  $\sim$0.65 with actual height; actual heights of most individuals in validation samples are within a few cm of the prediction. The variance captured for height is comparable to the
  estimated SNP heritability from GCTA (GREML) analysis, and seems
  to be close to its asymptotic value (i.e., as sample size goes to infinity), suggesting that we have
  captured most of the heritability for the SNPs used.
  Thus, our results resolve the common SNP portion of the ``missing heritability'' problem -- i.e., the gap between  prediction R-squared and SNP heritability. The $\sim$20k activated SNPs in our height predictor reveal the genetic architecture
  of human height, at least for common SNPs. Our
  primary dataset is the UK Biobank cohort, comprised of almost 500k
  individual genotypes with multiple phenotypes. We also use other datasets and SNPs found in earlier GWAS for out-of-sample validation of our results.

  \bigskip
\end{abstract}





\section{Introduction}

Recent estimates \cite{Yang2011} suggest that common SNPs account for significant heritability
of complex traits such as height, heel bone density, and educational attainment (EA). Large GWAS studies of these traits have identified many associated SNPs at genome-wide significance ($p < 5\times 10^{-8}$) \cite{Visscher2017, Marouli2017,Styrkarsdottir2008,Morris2017,Okbay2016}. However, the total variance accounted for by these SNPs is still a small fraction of the trait heritability and  of the proportion of variance that could be captured by regression on common SNPs as suggested by SNP heritability estimates \cite{delosCampos2015}.

The simplest hypothesis explaining this (so far) ``missing heritability'' is that previous studies have not had enough statistical power to identify most of the relevant SNPs, due to their small effect size, low minor-allele frequency (MAF), or both. In this letter, we provide evidence in support of this hypothesis by constructing genomic predictors capturing much of the estimated SNP heritability. We make use of a newly available large data set (the UK Biobank 500k genomes release) and new computational methods.

Association studies (GWAS) focus on reliable (high confidence) identification of associated SNPs. In contrast, genomic prediction based on whole genome regression methods~\cite{delosCampos2010} seeks to construct the most accurate predictor of phenotype, and tolerates possible inclusion of a small fraction of false-positive SNPs in the predictor set. The SNP heritability of the molecular markers used to build the predictor can be interpreted as an upper bound to the variance that could be captured by the predictor.

While identification of GWAS SNPs is accomplished by single SNP regression, construction of a best predictor is a global optimization problem in the high dimensional space of possible effect sizes of all SNPs. In this letter we use \Lone-penalized regression (LASSO or Compressed Sensing) to obtain our predictors. This method is particularly effective in cases where only a small subset of variables have non-zero effect on the predicted quantity (i.e., the effects vector is sparse, or approximately sparse). In earlier work \cite{Vattikuti2014} it was shown that matrices of human genomes are good compressed sensors, and that they are in the universality class of Gaussian random matrices. The \Lone algorithm exhibits phase transition behavior as the sample size and penalization parameter are varied; this behavior can be used to optimize the penalization as a function of sample size. Technical details are provided in the Methods section below.

Beyond the theoretical considerations given above, the practical outcome of our work is to significantly improve accuracy in genomic prediction of complex phenotypes. Using these predictors, one can, for example, reliably identify outliers in the population based on DNA alone. The activated SNPs in the predictors (i.e., those that have been assigned non-zero effect size by the LASSO algorithm) are likely to be associated with the phenotype, although they may not reach genome-wide significance in ordinary regression analysis. While there may be some contamination of false-positives among these SNPs, one can nevertheless infer properties of the overall genetic architecture of the trait (e.g., distribution of effect sizes with MAF).

\section{Data and Methods}

Our main dataset is the July 2017 release of nearly 500k UK Biobank genotypes and associated phenotypes~\cite{ukbb, Bycroft2017}. (See Supplement for more detailed description of data, quality control, algorithms, and computations.)

We compute an estimator $\vec{\beta}^*$ for the vector of
linear effects, $\vec{\beta}\in \R^p$, using \Lone-penalized regression
(LASSO)~\cite{Tibshirani94regressionshrinkage}. This corresponds to minimizing the objective
function below (phenotypes $\vec{y}$ are age and gender adjusted; both $\vec{y}$ and genotype values $X$ are standardized).
\begin{equation}
\vec{\beta}^* = \argmin_{\vec{\beta}\in\R^p} O_\lambda(\vec{y}, X; \vec{\beta}),\qquad
O_\lambda(\vec{y}, X; \vec{\beta}) = \frac{1}{2}\big\|\vec{y} - X\vec{\beta}\big\|^2
   + n\lambda \|\vec{\beta}\|_1,
\end{equation}
where $\lambda$ is a penalty (hyper-)parameter and the
\Lone norm is defined to be the sum of the absolute
values of the coefficients
\[
\|\vec{\beta}\|_1 = \sum_{j=1}^p |\beta_j|.
\]
The resulting effects vector $\vec{\beta}^*$ defines a linear predictive model which captures a large portion of the heritable genetic variance.

In our procedure, a first screening based on standard single marker regression is performed on the training set to reduce the set of candidate SNPs from 645,589 SNPs that passed QC (Supplement) to the top $p = 50$k and 100k by statistical significance.

\section{Results}

\begin{figure}[ht!]
\begin{center}
\includegraphics[height=4.5in,width=4.5in,keepaspectratio=true,angle=-90]{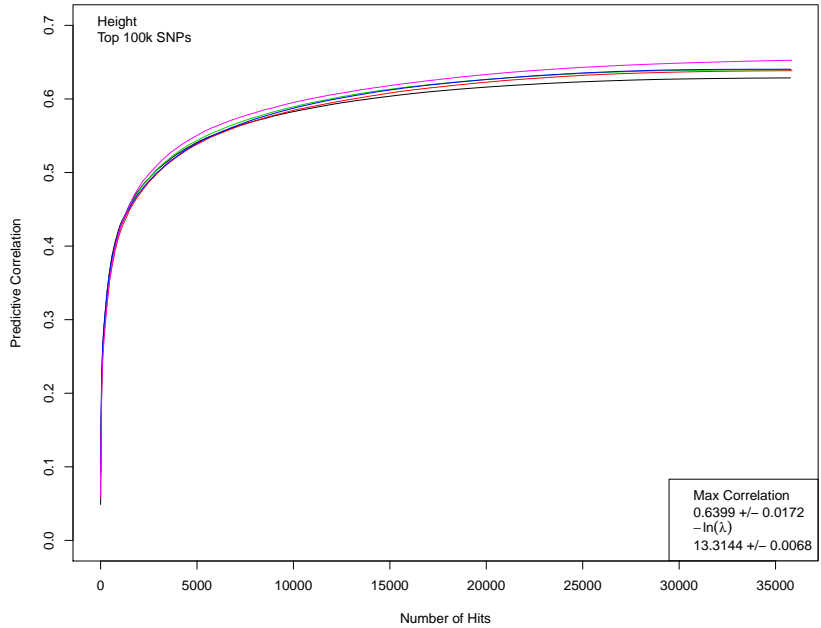}
 \caption{Correlation between actual and predicted heights as a function of the number of SNP hits activated in the predictor. While difficult to visually separate, each line represents the training of a predictor using 453k individuals. Correlation is computed on 5k individuals not used in training. The phase transition region (roughly, $10 < - \ln(\lambda) < 12$) corresponds to rapid growth in correlation on this graph, with number of hits growing from near 0 to over 5000. }
\label{fig:lassoscanhits}
\end{center}
\end{figure}

\begin{figure}[ht!]
\begin{center}
\includegraphics[height=4.5in,width=4.5in,keepaspectratio=true,angle=-90]{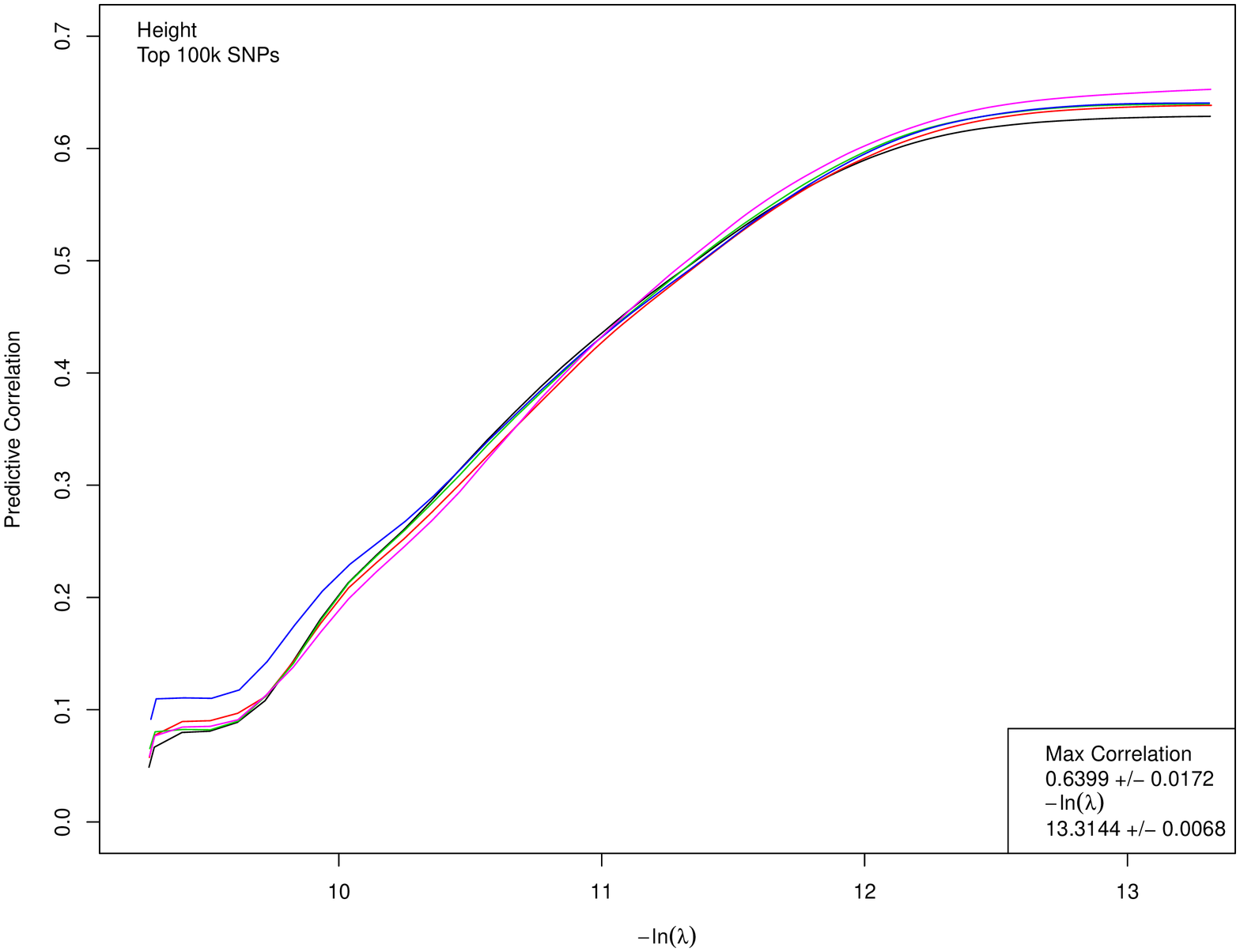}
 \caption{Correlation between actual and predicted heights as a function of \Lone penalization $\lambda$. Each line represents the training of a predictor using 453k individuals. Correlation is computed on 5k individuals not used in training.}
\label{fig:lassoscan}
\end{center}
\end{figure}

Figure (\ref{fig:lassoscanhits}) displays results from a typical LASSO run for height. 5 non-overlapping sets of 5k individuals each were held back from LASSO training using the top 100k candidate SNPs. For each value of the \Lone penalization $\lambda$ the resulting predictor $\vec{\beta}^*$ is applied to the genomes of the holdback sets and the correlation between predicted and actual height is computed. A phase transition (region of rapid variation in results) is expected and occurs at roughly $10 < - \ln(\lambda) < 12$. The penalization is reduced until the correlation is maximized. In Figure (\ref{fig:lassoscanhits}), the correlation is shown as a function of number of SNPs assigned non-zero effect sizes (i.e., activated) by LASSO. In the phase transition regime, where correlation rapidly increases, the number of activated SNPs grows rapidly from about zero to 7k. Each of the 5 colored curves in the figure corresponds to a training run on 453k individuals, with a different 5k held back (and slightly different training set) for each run. The phase transition is shown in terms of the penalization $- \ln(\lambda)$ in Figure (\ref{fig:lassoscan}).

\begin{figure}[ht!]
\begin{center}
\includegraphics[height=4.5in,width=4.5in,keepaspectratio=true,angle=-90]{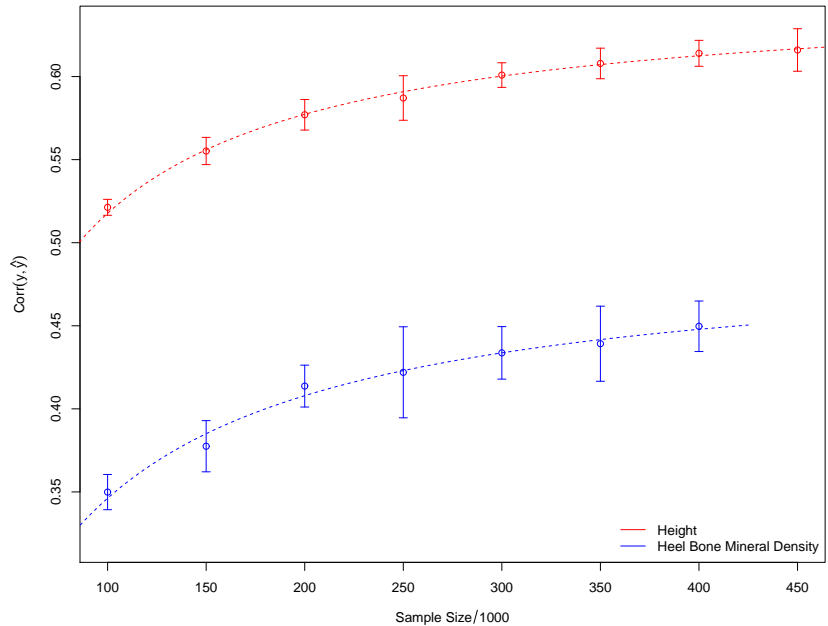}
 \caption{Correlation between predicted and actual height as number of individuals $n$ in training set is varied. $p = 50$k candidate SNPs used in optimization. Fit lines of the form Corr $\sim \frac{n}{n+b}$ are included to aid visualization.}
\label{fig:samplevary}
\end{center}
\end{figure}

\begin{figure}[ht!]
\begin{center}
\includegraphics[height=4.5in,width=4.5in,keepaspectratio=true]{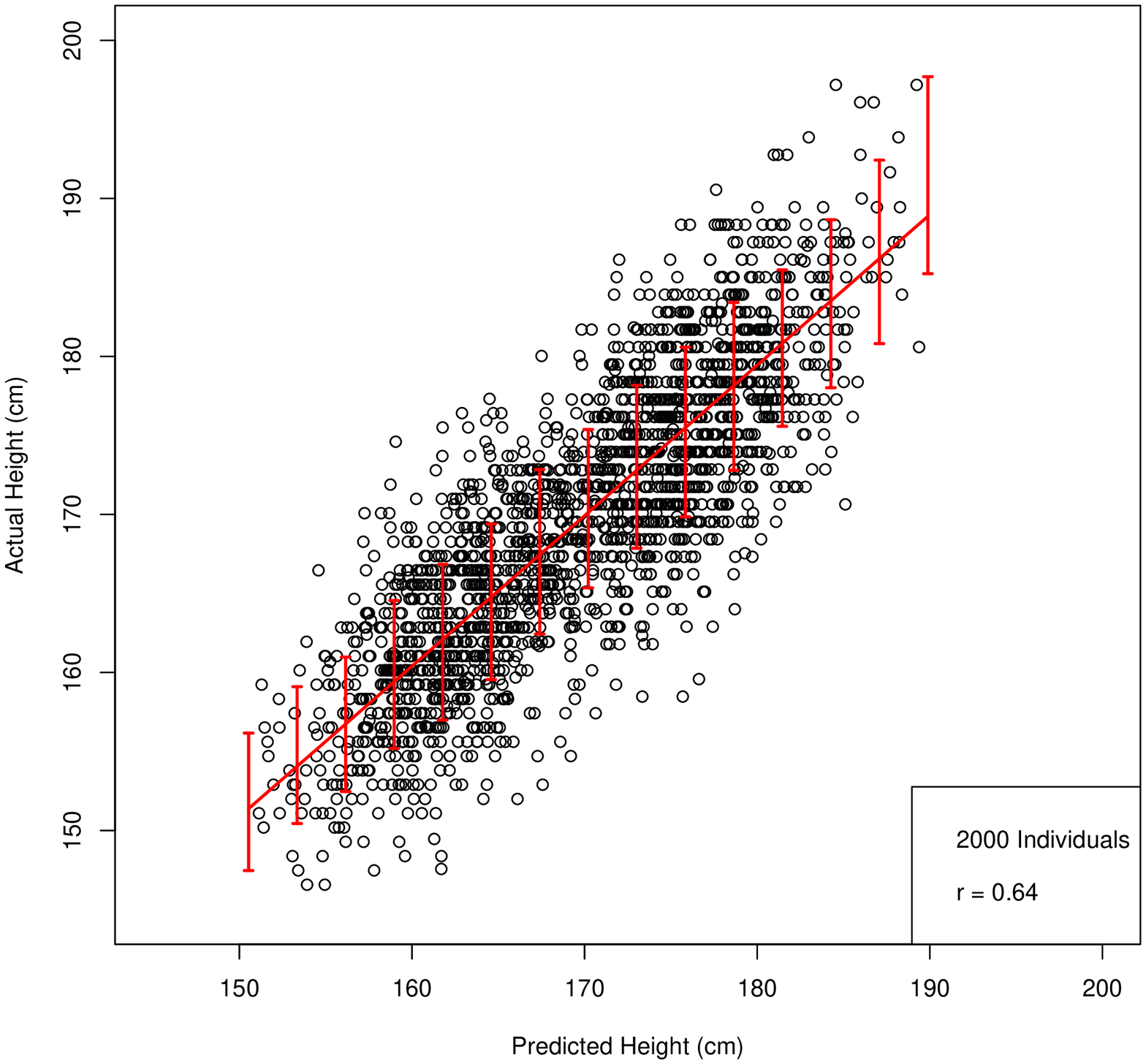}
 \caption{Actual height (cm) versus predicted height (cm) using 2000 randomly selected individuals held back from predictor optimization. (Roughly equal numbers of males and females; no corrections of actual height for age or gender). Error bars indicate $\pm 1$ SD range computed using larger validation set.}
\label{fig:hgtscatter}
\end{center}
\end{figure}

Figure (\ref{fig:samplevary}) shows the correlation between predicted and actual phenotypes in a validation set of 5000 individuals not used in the training optimization described in above - this is shown both for height and heel bone mineral density. The horizontal axis shows the number of individuals used in the training set and the error bars reflect 1 SD uncertainty estimated from five replications. The correlation obtained indicates convergence to an asymptotic value of somewhat less than 0.7 (corresponding to roughly 50 percent of total variance) for height, and perhaps 0.45 for heel bone mineral density. Figure (\ref{fig:hgtscatter}) shows a scatterplot (each point is an individual) of predicted and actual height for 2000 individuals (roughly equal numbers of males and females) not used in the training. The actual heights of most individuals are within about $3$ cm of the predicted value.

The corresponding result for Educational Attainment does not indicate any approach to a limiting value. Using all the data in the sample, we obtain maximum correlation of $\sim 0.3$, activating about 10k SNPs. Presumably, significantly more or higher quality data will be required to capture most of the SNP heritability of this trait.

The number of activated SNPs in the optimal predictors for height and bone density is roughly 20k. Increasing the number of candidate SNPs used from $p= 50$k to $p=100$k increased the maximum correlation of the predictors somewhat, but did not change the number of activated SNPs significantly.

We computed the GCTA heritability for the top 50k SNPs used, using randomly selected sets of 20k individuals. For height,  $h^2 = 0.5003 \pm 0.0209 ~~ (95\%)$ and heel bone density $h^2 = 0.4355 \pm 0.0226 ~~ (95\%)$. However, there has been debate in the literature over the statistical properties of GREML estimates of SNP heritability and it is not clear that standard estimation methods yield reasonably unbiased estimates even with large sample size~\cite{KrishnaKumar2015, delosCampos2015, Yang2016, Kumar2016, Gamazon2016}. Therefore, we suggest that GCTA estimates of SNP heritability should only be used as a rough guide. Perhaps the only way to determine the heritability of a trait over a specific set of genomic variants is to build the best possible predictor~\cite{Makowsky2011} (i.e., with, in principle, unlimited sample size $n$) to determine how much variance can be accounted for.

For height we tested out-of-sample validity by building a predictor model using SNPs whose state is available for both UKBB individuals (via imputation) and on Atherosclerosis Risk in Communities Study (ARIC)~\cite{aric} individuals (the latter is a US sample). This SNP set differs from the one used above, and is somewhat more restricted due to the different genotyping arrays used by UKBB and ARIC. Training was done on UKBB data and out-of-sample validity tested on ARIC data. A $\sim$5\% decrease in maximum correlation results from the restriction of SNPs and limitations of imputation: the correlation fell to $\sim$0.58 (from 0.61) while testing within the UKBB. On ARIC participants the correlation drops further by $\sim$7\%, with a maximum correlation of $\sim$0.54. Only this latter decrease in predictive power is really due to out-of-sample effects. It is plausible that if ARIC participants were genotyped on the same array as the UKBB training set there would only be a $\sim$7\% difference in predictor performance. An ARIC scatterplot analogous to Figure (\ref{fig:hgtscatter}) is shown in the Supplement. Most ARIC individuals have actual height within 4 cm or less of predicted height.

We also checked (see Supplement) that familial relationships in UKBB do not have an important impact on our results. LASSO training was done both on the full set of data and on a smaller data set where all first degree cousin or stronger relations were removed (kinship > 0.10). After filtering for kinship on the calls, this left 423,510 individuals for height and 382,727 individuals for heel bone density. This unrelated dataset was used for model training using random sets of 100k, 150k, ... , 400k individuals and there was no discernible difference in the results between using a training set drawn from the set of 423,510 kinship-filtered individuals and individuals from the unfiltered set.

\begin{figure}[ht!]
\begin{center}
\includegraphics[height=4.5in,width=4.5in,keepaspectratio=true,angle=-90]{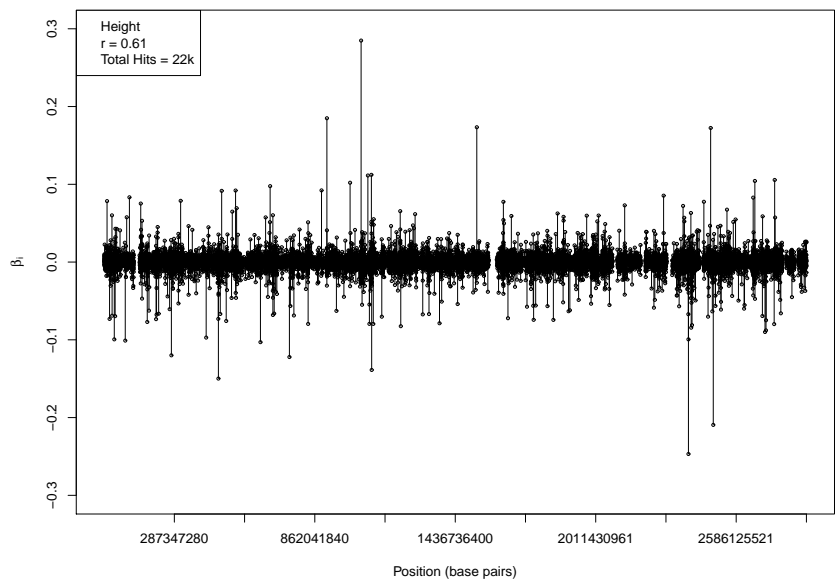}
 \caption{Effect size (minor allele) for each activated SNP in a predictor model. The horizontal axis represents the SNP position in the genome, if chromosomes (1-22) were connected end to end to form a continuous linear region. Activated SNPs are distributed roughly uniformly throughout the genome.}
\label{fig:betamanhat}
\end{center}
\end{figure}

The genetic architecture of a height model is displayed in Figure (\ref{fig:betamanhat}), which shows the effect size (minor allele) for each activated SNP. The horizontal axis represents the SNP position in the genome, if each chromosome (1-22) were laid end to end to form a continuous linear region. The specific height predictor from which these SNPs are taken was built from 50k candidate SNPs and achieves a correlation between actual and predicted height of $\sim$0.61. The activated SNPs seem to be uniformly distributed across the genome.

There is significant overlap between regions of the genome near previously known SNPs and regions identified by our algorithm (Supplement). However, our activated SNPs are roughly uniformly distributed over the entire genome, and number in the many thousands for each trait. This means that many of our SNPs, including some of those that account for the most variance, are in regions not previously identified by earlier GWAS.


\section{Discussion}

Until recently most work with large genomic datasets has focused on finding {\it associations} between markers (e.g., SNPs) and phenotype~\cite{Makowsky2011}. In contrast, we focused on optimal {\it prediction} of phenotype from available data. We show that much of the expected heritability from common SNPs can be captured, even for complex traits affected by thousands of variants. Recent studies using data from the interim release of the UKBB reported prediction correlations of about 0.5 for human height using roughly 100K individuals in the training~\cite{hwasoon}. These studies forecast further improvement of prediction accuracy with increased sample size, which have been confirmed here.

We are optimistic that, given enough data and high quality phenotypes, results similar to those for height might be obtained for other quantitative traits, such as cognitive ability or specific disease risk. There are numerous disease conditions with heritability in the 0.5 range, such as Alzheimer's, Type I Diabetes, Obesity, Ovarian Cancer, Schizophrenia, etc~\cite{heritabilityestimates}. Even if the heritable risk for these conditions is controlled by thousands of genetic variants, our work suggests that effective predictors might be obtainable (i.e., comparable to the height predictor in Figure (\ref{fig:hgtscatter})). This would allow identification of individuals at high risk from genotypes alone. The public health benefits are potentially enormous.

We can roughly estimate the amount of case-control data required to capture most of the variance in disease risk. For a quantitative trait (e.g., height) with $h^2 \sim 0.5$, our simulations \cite{Vattikuti2014} predict that the phase transition in LASSO performance occurs at $n \sim 30 s$ where $n$ is the number of individuals in the sample and $s$ is the sparsity of the trait (i.e., number of variants with non-zero effect sizes). For case-control data, we find $n \sim 100 s$ (where $n$ means number of cases with equal number controls) is sufficient. Thus, using our methods, analysis of $\sim 100$k cases together with a similar number of controls might allow good prediction of highly heritable disease risk, even if the genetic architecture is complex and depends on a thousand or more genetic variants.

\paragraph{Acknowledgments}

LL, SA, and SH acknowledge support from the Office of the Vice-President for Research at MSU. The authors are grateful for useful correspondence and discussion with Alexander Grueneberg and Hwasoon Kim. We also acknowledge support from the NIH Grants R01GM099992 and R01GM101219, and NSF Grant IOS-1444543, subaward UFDSP00010707.

\newpage
\appendix

\section{SUPPLEMENT: Methods}

\subsection{UKBB Dataset QC}

In July 2017, the UK Biobank~\cite{ukbb, Bycroft2017} released a set of 488,377 genotyped individuals which were genotyped using two Affymetrix platforms---approximately 50,000 samples on the UK BiLEVE Axiom array and the remainder on the UK Biobank Axiom array. The initial genotype information was collected for 488,377 individuals for 805,426 SNPs and then subsequently imputed. Quality Control was done on the un-imputed data by removing SNPs which had missing call rates over 3\%, removing individuals which had missing call rates over 10\% and, so as not to deal with very rare variants, removing SNPs which had minor frequencies below 0.1\%. The resulting genetic data contained 645,589 SNPs and 488,371 individuals. This set was then further filtered for self-reported caucasians for whom the necessary phenotype measurements were available: for height, the number of remaining individuals was 457,484; for heel bone mineral density there were 413,444 individuals; and for educational attainment, there were 455,637 individuals.

The imputed data set was generated using the set of 805,426 raw markers using the Haplotype Reference Consortium and UK10K haplotype resources. After imputation and initial QC, there were a total of 92,060,613 SNPs and 487,411 individuals. From this imputed data, further quality control was performed using Plink version 1.9 by excluding SNPs and samples which had missing call rates exceeding 3\% and also removing snps with minor allele frequency below 0.1\%. For out-of-sample validation of height, we extracted SNPs which survived the prior quality control measures, and are also present in a second dataset from the Atherosclerosis Risk in Communities Study (ARIC)~\cite{aric}. This resulted in a total of 632,155 SNPs and 464,192 samples. All quality control steps, except those performed by the UK Biobank involving the imputation, were performed using version 1.9 of the Plink software~\cite{Purcell2007}.


\subsection{Confounding variables: age, sex and family structure}

All traits for self-identified Caucasians were adjusted on the basis of age and sex. The phenotypes for self-reported Caucasians were adjusted by z-scoring the phenotypes amongst all individuals of the same sex. To correct for the effects of societal changes, a univariate linear regression was performed on z-scored phenotypes using year-of-birth as the dependent variable. The adjusted phenotype was set equal to the residual of the z-scored phenotype and the regression line. Before making these corrections, it was shown that the mean phenotypic value was indeed increasing with year-of-birth---this was seen in all three phenotypes: height, heel bone mineral density and educational attainment.

Relatedness calculations were provided with the UKBB dataset in order to account for family structure and cryptic relatedness. There were 107,163 familial relationships identified amongst UKBB participants which were at the level of third cousins or higher and, due to the large number of relationships, filtering out these individuals results in a nontrivial decrease in the size of data available for model selection. To investigate the relevance of this issue, LASSO training was done both on the full set of data and on a smaller data set where all first degree cousin or stronger relations were removed (kinship > 0.10). After filtering for kinship on the calls, this left 423,510 individuals for height and 382,727 individuals for heel bone density. This unrelated dataset was used for model training using random sets of 100, 150, ... , 400 thousand individuals and there was no discernible difference in the results between using a training set drawn from the set of 423,510 kinship-filtered individuals and individuals from the unfiltered set. Therefore we do not believe that the familial relationships have an important impact on our results.

\subsection{\Lone-penalized regression}

Consider the regression problem in generality. We
have $n$ observations of the phenotype, $y_I$, with $I=1,\dots, n$ as
the vector $\vec{y}$. The genotype data is encoded in the $n\times p$
design matrix $X_{Ij}$ with $j=1,\dots, p$. The $X_{Ij}$ is the number
of copies of the most frequent minor allele of the $j$th SNP for the
$I$th person, and thus takes values $0$, $1$, or $2$. Missing values
are mean-imputed.

We use a standard linear model for the dependence of $y$
on the SNP data $x_{j}$. That is, we assume a relationship of the
form
\begin{equation}
  y_I = y_0 + \hat{\vec{\beta}}\cdot\vec{x}_I + e_I,
\end{equation}
where the errors, $e_I$, are assumed to be (identically and
independent) normally distributed with unknown variance
$\sigma_e$. The errors, $e_I$, receive contributions from potential
environmental effects, gene--gene nonlinear effects, and
gene--environment nonlinear effects. For discussion of methods to
recover nonlinear effects, see \cite{Ho2015}.

We compute an estimator $\vec{\beta}^*$ for the vector of
linear effects, $\hat{\vec{\beta}}\in \R^p$, using \Lone-penalized regression
(LASSO)~\cite{Tibshirani94regressionshrinkage}. This corresponds to minimizing the objective
function (after standardizing $\vec{y}$ and $X$)
\begin{equation}\label{eq:lasso-def}
\vec{\beta}^* = \argmin_{\vec{\beta}\in\R^p} O_\lambda(\vec{y}, X; \vec{\beta}),\qquad
O_\lambda(\vec{y}, X; \vec{\beta}) = \frac{1}{2}\big\|\vec{y} - X\vec{\beta}\big\|^2
   + n\lambda \|\vec{\beta}\|_1,
\end{equation}
where $\lambda$ is a penalty (hyper-)parameter and the
\Lone norm is defined to be the sum of the absolute
values of the coefficients
\[
\|\vec{\beta}\|_1 = \sum_{j=1}^p |\beta_j|.
\]
(We use $\|\cdot\|$ with no subscript to denote the standard L\textsubscript{2} norm.)
The extra factor of $n$ in the second term is a convention that
factors out the explicit sample size scaling of $n$. The squaring in
the first term is (implicitly) of the L\textsubscript{2} norm of the
residual.

The first term is the standard ordinary least-squares (OLS) loss
function. The effect of the second term is to regularize the
regression problem by favoring sparse solutions with the nonzero
coefficients shrunk toward $0$. This seems appropriate for genomic
problems, since we expect that for any given phenotype most SNPs have
no effect. Biasing the nonzero coefficients toward $0$ reduces
variance and improves the expected fit for small sample size.

Even for
$n \ll p$, LASSO can obtain an accurate $\vec{\beta}$ under the right conditions: the effects vector must be sparse and the heritability of the trait must be sufficiently high (equivalently, the amount of noise variance is bounded).
For fixed $\sigma_e^2$ and sparse effects vector, there is a critical sample size $n^*$ (depending
on $\sigma_e$ and the sparsity of the trait) above which one expects
to get good recovery of $\vec{\beta}$ in terms of the
L\textsubscript{2} error. A phase transition at $n \sim n^*$ has been
demonstrated numerically for real and simulated genomic data
in~\cite{Vattikuti2014}.

For our specific calculations we follow the following cross-validation
procedure:
\begin{enumerate}
\item Break the data into training sets, and smaller test and validation sets.
\item Perform a standard GWAS on the training sets, and rank the SNPs by $p$-value.
\item To ease the computational burden, restrict the calculation to a
  fixed number of lowest $p$-value SNPs on each training set.
  Replace any missing SNP values by the SNP-mean for the
  training data.
\item Perform LASSO on the standardized training data, scanning a range of values
  for the penalty $\lambda$ that passes through the phase transition region of rapid variation in results.
\item Choose the $\lambda$ that has the maximum correlation on the test set, which was held back from training.
\item Finally, evaluate performance of optimal predictor $\beta^*$ on validation sets.
\end{enumerate}

\subsection{Coordinate Descent}

Most algorithms for minimizing the objective
function~\eqref{eq:lasso-def} use (some variation of) coordinate
descent~\cite{Friedman2007, Friedman2010}.\footnote{We use a custom
  implementation in~Julia~\cite{julia} using safe screening
  ideas~\cite{safe,sasvi,gap,rfne}.} The basic form of the algorithm
is as follows. Proceeding from an initial guess $\vec{\beta}_0$, we
cycle through the $p$ ``coordinates'' sequentially, minimizing $O$
with respect to each $\beta_j$ (holding others fixed). To that end, note that
\begin{equation}
\prd{O}{\beta_j} = n \left[\vev{x_j^2}\beta_j + \sum_{k\neq j} \vev{x_j x_k}\beta_k - \vev{x_j y} + \lambda\sign(\beta_j)\right] = 0.
\end{equation}
Thus, the updated coefficient should satisfy
\begin{equation}
\beta_j^* = \frac{1}{\vev{x_j^2}}\left[\vev{x_j y} - \sum_{k\neq j} \vev{x_j x_k}\beta_k - \lambda\sign(\beta_j^*)\right].
\end{equation}
To solve for $\beta_j^*$, one should determine the $\lambda = 0$
solution. If it is positive (negative), then guess that
$\sign(\beta_j^*)$ should be positive (negative) and subtract (add)
the $\lambda$ term. If the sign flips, then the solution is spurious,
and the optimal solution is at $\beta_j^* = 0$. (To see this note that
for $\beta_j^* = 0^+$ the derivative is positive, and for
$\beta_j^* = 0^-$ the derivative is negative.)

Introduce the ``soft thresholding function''
\begin{equation}
S(z, \gamma) = \sign(z)\max(|z|-\gamma, 0).
\end{equation}
Then, the update for the $j$th component of $\vec{\beta}$ is
\begin{equation}
\beta_j^* = \frac{1}{\vev{x_j^2}}
  S\left(\vev{x_j y} - \sum_{k\neq j} \vev{x_j x_k}\beta_k,\,\lambda\right).
\end{equation}
The basic Coordinate descent algorithm is as shown in Alg.~\ref{alg:coord-desc}.

\begin{algorithm}[h]
\DontPrintSemicolon
\KwData{$X_{jI}$ and  $y_I$ with $j=1,\dots, p$ and $I=1,\dots, n$}
\KwIn{Penalty parameter $\lambda$, tolerance $\epsilon$, and (optionally) initial guess $\vec{\beta}_{0}$}
\KwOut{$\vec{\beta}$ solving LASSO optimization problem within convergence tolerance $\epsilon$}
$\vec{\beta}\gets \vec{\beta}_0$\;
\Repeat{$(\vec{\beta} - \vec{\beta}_0)^2 < \epsilon^2$}
{
$\vec{\beta}_0\gets \vec{\beta}$\;
\For{$j$ \textbf{in} $\{1,\dots, p\}$ }
{
$\beta_j\gets \frac{1}{\vev{x_j^2}}
  S\left(\vev{x_j y} - \sum_{k\neq j} \vev{x_j x_k}\beta_k,\,\lambda\right)$\;
}
}
\Return{$\vec{\beta}$}\;
\caption{Basic coordinate descent algorithm for LASSO.}
\label{alg:coord-desc}
\end{algorithm}

\subsection{Out-of-sample Validation}

Model (i.e., predictor) construction was performed by implementing LASSO on the UK Biobank
 data. In order to validate models and check against overtraining, a
second dataset is needed in order to test the results.
We 1) withheld a small subset of UKBB individuals from the
initial training for in-sample validation, and 2) applied the model
to individuals from a completely different dataset (ARIC) for out-of-sample
validation. In-sample validation was done by withholding a
predetermined number of randomly selected individuals from the UK
Biobank data before p-value cuts were applied to SNPs. The remaining
individuals were used for LASSO training and the resulting model
was applied to the individuals initially held back to check in-sample validity.

\begin{figure}[ht!]
\begin{center}
\includegraphics[height=4.5in,width=4.5in,keepaspectratio=true]{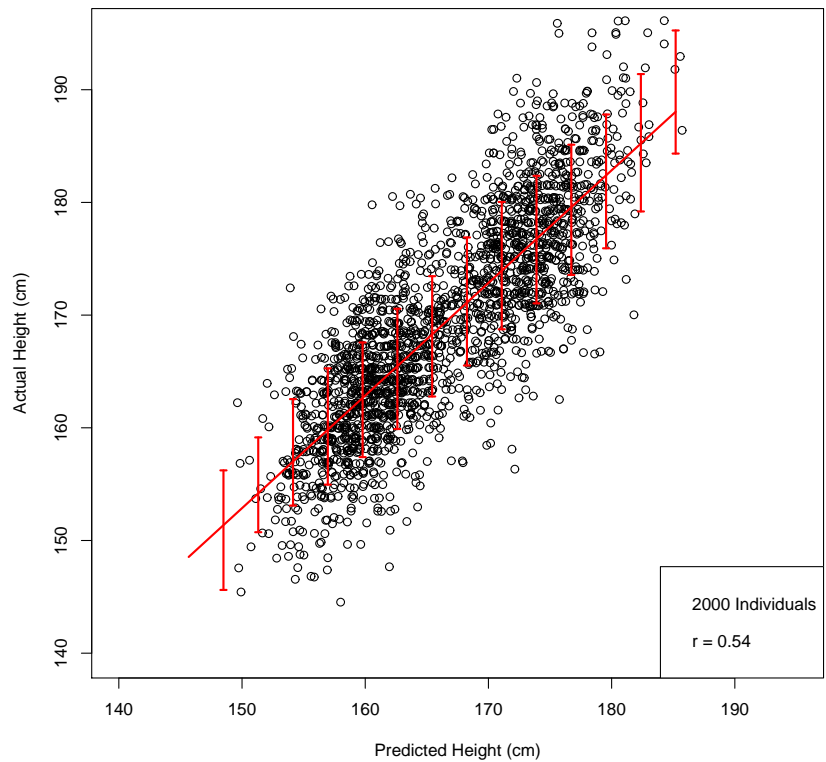}
 \caption{Actual height (cm) versus predicted height (cm) using 2000 randomly selected individuals (roughly equal numbers of M and F; no corrections for age or gender) from the ARIC dataset. Error bars indicate $\pm 1$ SD range computed using larger validation set.}
\label{fig:aricscat}
\end{center}
\end{figure}

Out-of-sample validation is similar, except that we used a set of common SNPs
for which state values can be imputed on the UKBB individuals and are also known for
ARIC individuals. Initial training of the model was performed using UKBB individuals, but
its validity was then tested on the ARIC data. Results using the un-imputed dataset reached
correlation of $\sim$0.61 when testing within the UKBB. After selecting SNPs in common with ARIC, the correlation
fell to $\sim$0.58 while testing within the UKBB and achieved a correlation of $\sim$0.54 on ARIC participants. The
ARIC results are shown in Figure (\ref{fig:aricscat}). Actual heights of most individuals in the ARIC validation set are within 4 cm or less of the predicted height.

The ARIC dataset~\cite{aric} was composed of 12,772 caucasian and African-American individuals who were genotyped on the Affymetrix 6.0 chip with 841,820 SNPs. This was filtered to keep only caucasian individuals and SNPs with MAF larger than 1\% and missing call rates below 5\% with a final sample size of 9618 individuals with 705,956 SNPs. After filtering to only SNPs which were in common with the UKBB imputed data, the number of SNPs was reduced to 632,155.


\begin{figure}[ht!]
\begin{center}
\includegraphics[height=4.5in,width=4.5in,keepaspectratio=true,angle=-90]{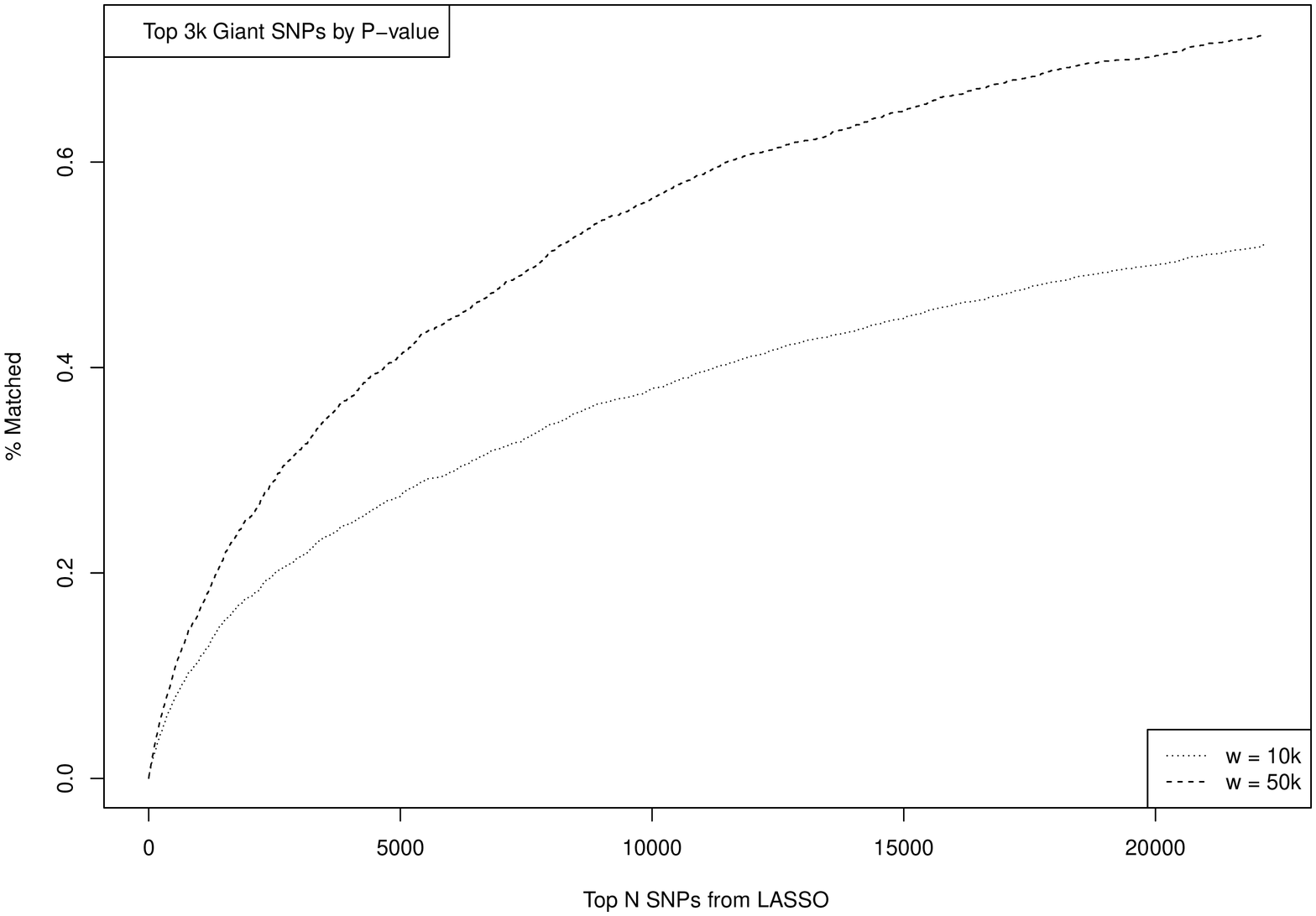}
 \caption{Matching between top SNPs activated in predictor model ordered by variance accounted for (x axis) and SNPs identified previously by GIANT GWAS (height). Percent of previously known SNPs matched is shown on y axis. Matching window size $w$ given in bp.}
\label{fig:giantcomp}
\end{center}
\end{figure}

\begin{figure}[ht!]
\begin{center}
\includegraphics[height=4.5in,width=4.5in,keepaspectratio=true,angle=-90]{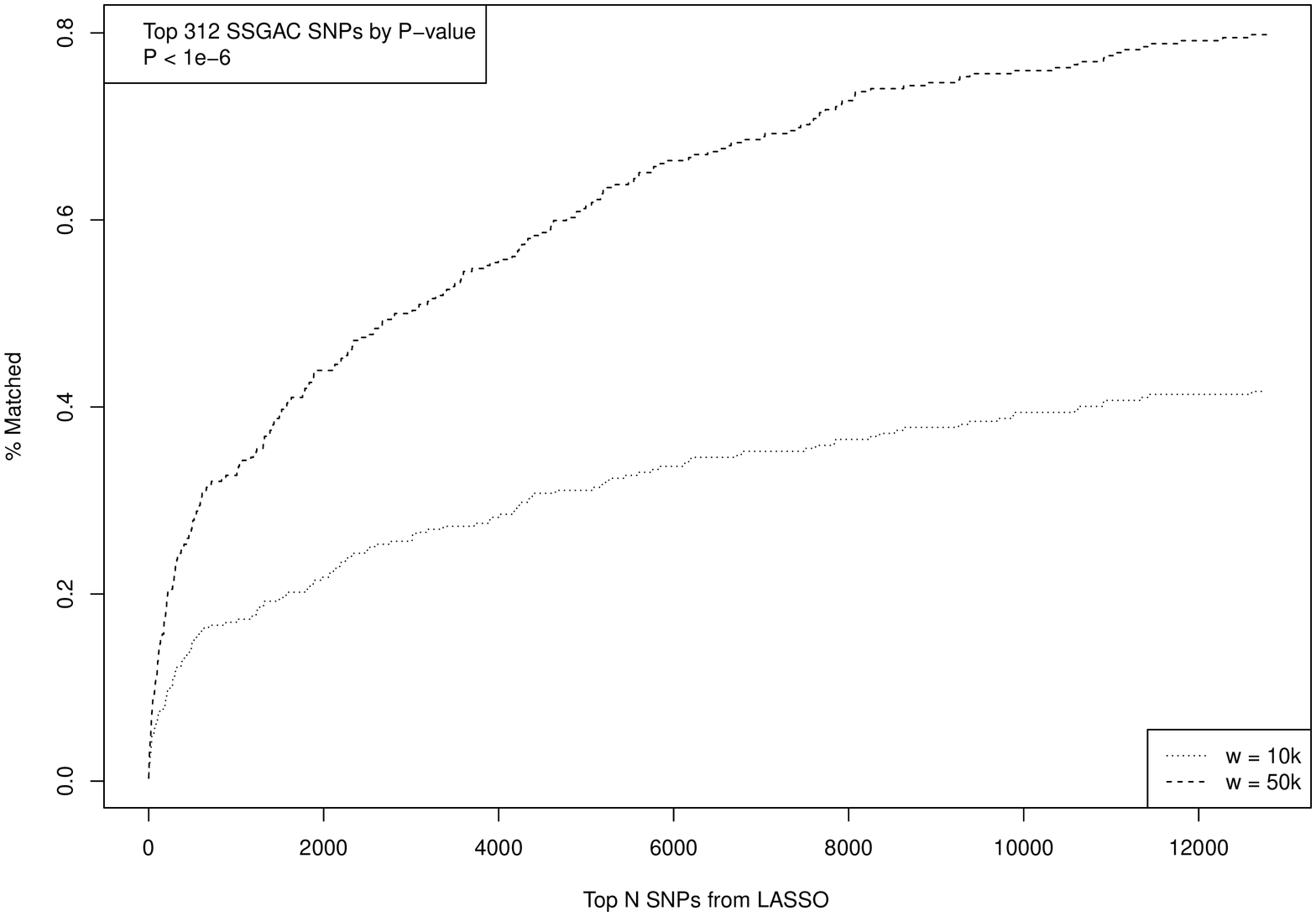}
 \caption{Matching between top SNPs activated in predictor model ordered by variance accounted for (x axis) and SNPs identified previously by SSGAC GWAS (Educational Attainment). Percent of previously known SNPs matched is shown on y axis. Matching window size $w$ given in bp.}
\label{fig:ssgaccomp}
\end{center}
\end{figure}

\begin{figure}[ht!]
\begin{center}
\includegraphics[height=4.5in,width=4.5in,keepaspectratio=true,angle=-90]{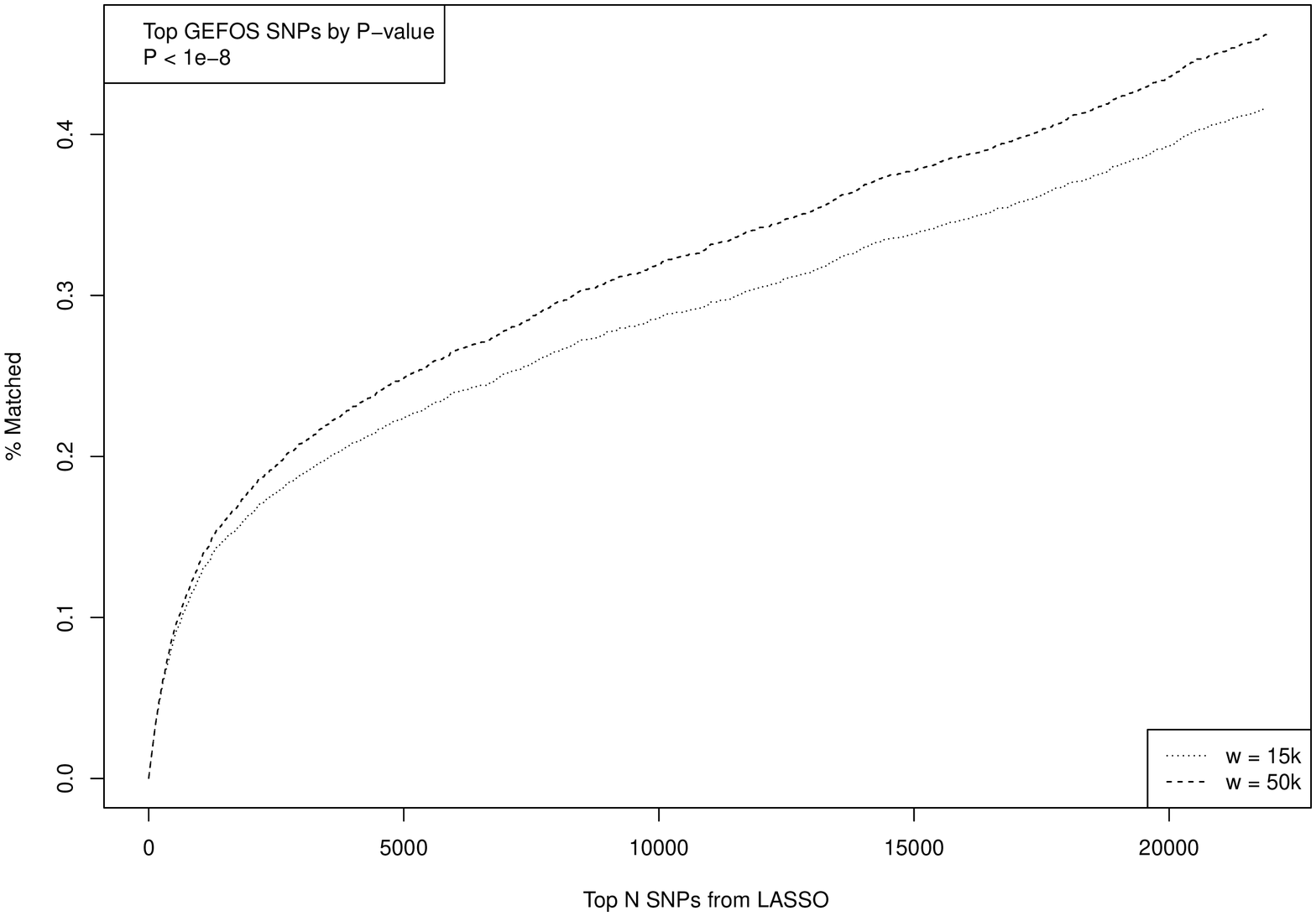}
 \caption{Matching between top SNPs activated in predictor model ordered by variance accounted for (x axis) and SNPs identified previously by GEFOS GWAS (Heel Bone Density). Percent of previously known SNPs matched is shown on y axis. Matching window size $w$ given in bp.}
\label{fig:gefoscomp}
\end{center}
\end{figure}

We compare our activated predictor SNPs to known hits from GWAS collaborations studying the same phenotypes \cite{Marouli2017,Styrkarsdottir2008,Morris2017,Okbay2016}. Specifically we compare our results for height with those of the GIANT collaboration, for educational attainment with SSGAC, and for Bone Density with GEFOS.
We ordered activated SNPs (i.e., those assigned non-zero effect size $\beta$ by the LASSO algorithm) by variance explained ($V_i = 2 \nu (1-\nu) \beta_i^2$ where $\nu$ is the minor allele frequency), then scanned down this list and looked for a proxy match by distance in the corresponding dataset. For GIANT, we took the results published online and extracted the top 3000 hits ordered by p-value. For SSGAC, we used the published results and kept SNPs with $p <  10^{-6}$ - a total of 316. For GEFOS, we kept all SNPs with $p < 10^{-8}$ and then coarse grained SNPs within blocks of 10k base pairs, resulting in 3901 regions. These results are displayed in Figures (\ref{fig:giantcomp}), (\ref{fig:ssgaccomp}), (\ref{fig:gefoscomp}). They show significant overlap between regions of the genome near previously known SNPs and regions identified by our algorithm. However, our activated SNPs are roughly uniformly distributed over the entire genome, and number in the many thousands for each trait. This means that many of our SNPs, including some of those that account for the most variance, are in regions not previously identified by earlier GWAS.

\begin{figure}[ht!]
\begin{center}
\includegraphics[height=4.5in,width=4.5in,keepaspectratio=true,angle=-90]{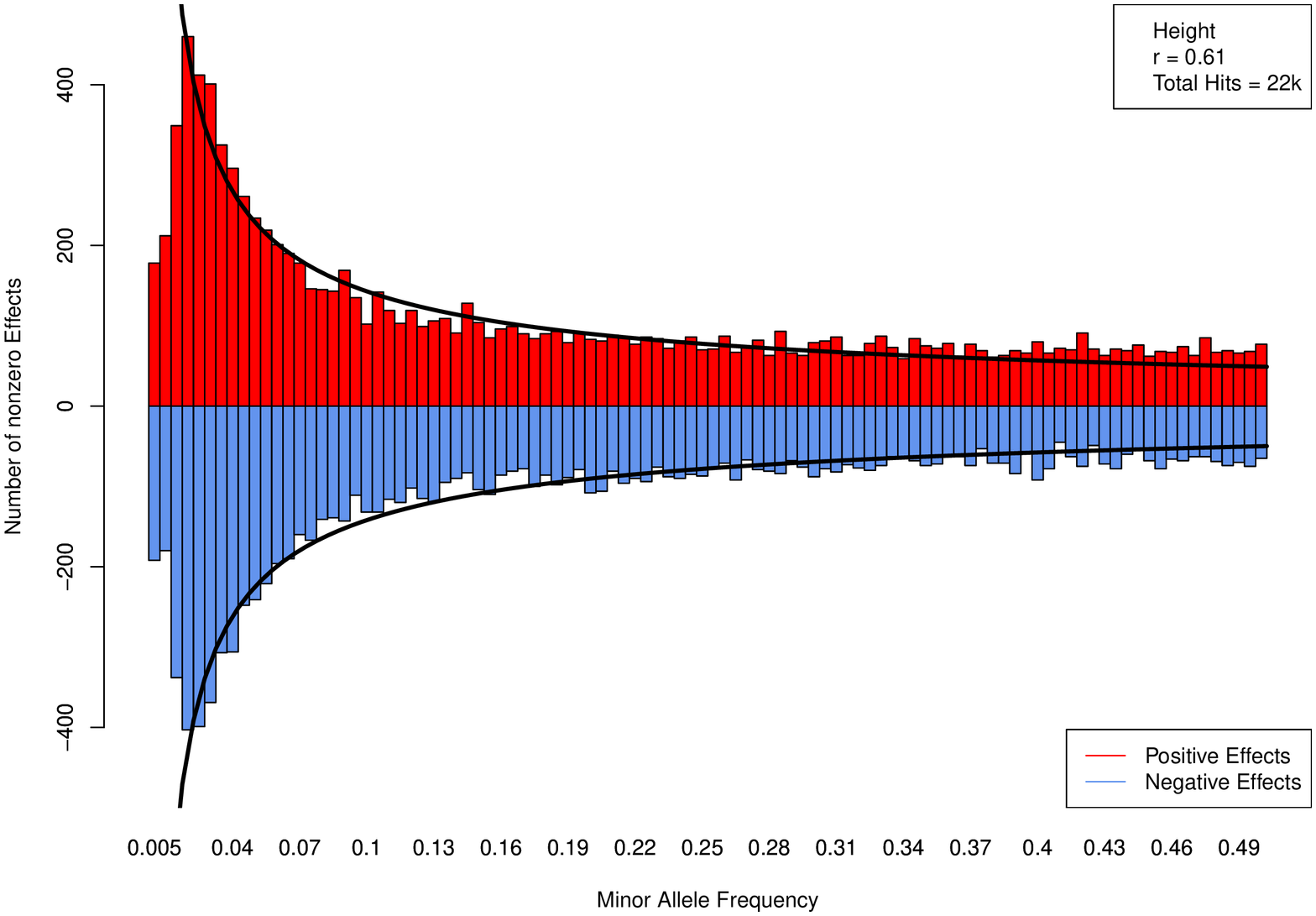}
 \caption{Number of SNPs with positive (red) and negative (blue) minor allele effect sizes. Curves are constructed by fitting a power law in MAF.}
\label{fig:betafx}
\end{center}
\end{figure}

Figure (\ref{fig:betafx}) shows number of activated SNPs by {\it sign of effect of the minor allele} and minor allele frequency (MAF). The height of each bar represents the number of $+$ or $-$ SNPs in a MAF bin of width $0.005$. The specific height predictor from which these SNPs are taken was built from 50k candidate SNPs and achieves a correlation between actual and predicted height of $\sim$0.61. The curves, which are meant to aid visualization, are constructed by fitting a power law $n(\nu) = a {\nu}^{-b}$ to the range $\nu \in (0.025,0.3)$ where $\nu$ is MAF and $n(\nu)$ is the number of nonzero effects. We exclude the smallest values of MAF because of incomplete discovery of SNPs in that region. The $\pm$ distributions are nearly symmetrical ($a_+ = 31.07 , b_+ = 0.6553$; $a_- = 31.96; b_- = 0.6404$), even at very small MAF. There does not appear to be a statistically significant deviation from random assignment of signs -- the minor allele of an activated SNP is just equally likely to increase or decrease height.

\printbibliography
\end{document}